\documentclass[11pt,draftcls,peerreview,onecolumn]{IEEEtran}

\usepackage{cite}
\usepackage{subfigure}
\usepackage{graphicx}
\usepackage{url}
\usepackage{amsmath}
\usepackage{amssymb}
\newtheorem{theorem}{Theorem}

\newcommand{\pibold}{\mbox{\boldmath$\pi$}}
\newcommand{\lamNbold}{\mbox{\boldmath$\lambda_N$}}

\begin{document}

\title{Random Access Broadcast: Stability and Throughput Analysis}

\author{Brooke~Shrader,~\IEEEmembership{Student Member, IEEE}
        and~Anthony~Ephremides,~\IEEEmembership{Fellow,~IEEE}% <-this % stops a space
%%\thanks{Manuscript received August 31, 2005; revised January 15, 2006.
%%        This work was supported by...}% <-this % stops a space
\thanks{This work is supported in part by the Office of
Naval Research through grant N000140610065 and by the Department of
Defense under MURI grant S0176941. Prepared through collaborative
participation in the Communications and Networks Consortium
sponsored by the U. S. Army Research Laboratory under The
Collaborative Technology Alliance Program, Cooperative Agreement
DAAD19-01-2-0011. The U. S. Government is authorized to reproduce
and distribute reprints for Government purposes notwithstanding any
copyright notation thereon.
B. Shrader and A. Ephremides are with
the University of Maryland, College Park. The material in this paper
was presented in part at the 2005 International Symposium on
Information Theory~(ISIT) in Adelaide, Australia.}}

\markboth{Submitted to IEEE Transactions on Information Theory,
September 2006. Revised April 2007.}{Shell \MakeLowercase{\textit{et
al.}}: Random Access Broadcast: Stability and Throughput Analysis}

% If you want to put a publisher's ID mark on the page
% (can leave text blank if you just want to see how the
% text height on the first page will be reduced by IEEE)
%\pubid{0000--0000/00\$00.00~\copyright~2006 IEEE}
\maketitle

\begin{abstract}

A wireless network in which packets are broadcast to a group of
receivers through use of a random access protocol is considered in
this work.  The relation to previous work on networks of interacting
queues is discussed and subsequently, the stability and throughput
regions of the system are analyzed and presented. A simple network
of two source nodes and two destination nodes is considered first.
The broadcast service process is analyzed assuming a channel that
allows for packet capture and multipacket reception. In this small
network, the stability and throughput regions are observed to
coincide. The same problem for a network with $N$ sources and $M$
destinations is considered next. The channel model is simplified in
that multipacket reception is no longer permitted. Bounds on the
stability region are developed using the concept of stability rank
and the throughput region of the system is compared to the bounds.
Our results show that as the number of destination nodes increases,
the stability and throughput regions diminish. Additionally, a
previous conjecture that the stability and throughput regions
coincide for a network of arbitrarily many sources is supported for
a broadcast scenario by the results presented in this work.

\end{abstract}


% Will be ignored for other modes.
\IEEEpeerreviewmaketitle

%====================
\section{Introduction}
%====================

%\PARstart{T}{his} is the way to make the first letter big
%\hfill November 18, 2002
% needed in second column of first page if using \pubid
%\pubidadjcol

% Reminder: the "draftcls" or "draftclsnofoot", not "draft", class option
% should be used if it is desired that the figures are to be displayed while
% in draft mode.

The stability and throughput of finite-user random access systems
for {\it unicast} transmission have been studied extensively. The
throughput analysis is included in some of Abramson's early work on
the topic \cite{Abramson77}, while the stability problem was first
introduced by Tsybakov and Mikhailov \cite{TsybakovMikhailov}. The
finite-user stability problem proves to be much more difficult than
the throughput problem, and as such, the history of the stability
problem is more rich and interesting. In \cite{TsybakovMikhailov},
sufficient conditions for ergodicity were found using the transition
probabilities of the Markov chain corresponding to queue lengths.
Later work \cite{RaoAE88} described the use of stochastic dominance
as a means of characterizing the stability region. Additionally,
stability conditions based on the joint queue statistics were
provided in \cite{Szpa94}. Despite these and other attempts, the
stability region for arbitrarily many sources $N$ remains unsolved.
In \cite{LuoAE99} the concept of stability rank was introduced and
provided the tightest known bounds to the stability region for $N$
sources. The exact stability region for $N$ sources and an arrival
process which is correlated among the sources was obtained in
\cite{Anantharam}.

Most works, including all of those mentioned above, study random
access under the collision channel model, in which transmission by
more than one source results in failed reception of all packets.
Some recent works have incorporated the probabilistic nature of
reception and the possibility of multipacket reception (MPR) into
the channel model. The stability of infinite-user random access with
MPR was first examined in \cite{GhezEtAl88}. More recently, the
finite-user problem was examined in \cite{NawareEtAl03} and it was
shown that the possibility of MPR results in an increase in the
stable throughput of the system. The benefit to stability was so
dramatic that for channels with sufficiently strong reception
capabilities, random access was shown to outperform time division
multiple access (TDMA) schemes. This result provides motivation for
a renewed interest in random access.

In this correspondence we introduce {\it multiple destination nodes}
and broadcast transmission into the network and study the resulting
stability and throughput performance of random access. The
introduction of multiple destinations is a necessary first step in
understanding the behavior of ad hoc and multihop networks, where
random access presents an advantage over TDMA due to its distributed
nature. In particular, we analyze the performance of a random access
{\it broadcast} system in which a source node sends a common packet
to all destination nodes. Broadcast transmission is useful for
control of the network (ie, initialization, route discovery, timing
synchronization) and for a number of applications.

%======================================================
\section{Model and formulation} \label{section:Model}
%======================================================

\begin{figure}
\centering
\includegraphics[width=2.5in]{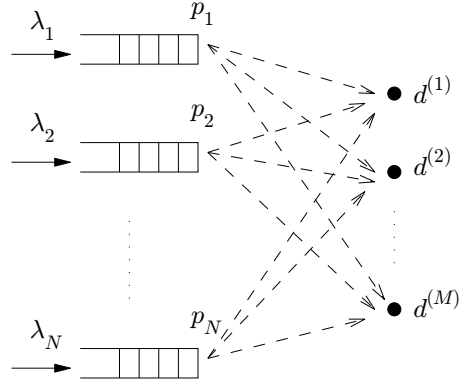}
\caption{The broadcast scenario we study in this work: $N$ source nodes transmit to $M$ destination nodes.}
\label{fig:multicast_NsourceMdest}
\end{figure}

Consider a system ${\cal S}$ consisting of $N$ source nodes, $s_1,
s_2, \hdots, s_N$, and $M$ destination nodes $d^{(1)}, d^{(2)},
\hdots, d^{(M)}$. Packets arrive to $s_n$ according to a Bernoulli
process with rate $\lambda_n$, $n=1,2,\hdots,N$ packets per slot.
The arrival process is independent from source to source and
independent, identically distributed (i.i.d) over slots. Packets
that are not immediately transmitted are stored in an infinite
buffer maintained at each source. All source nodes compete in a
random fashion for access to the channel in order to transmit a
packet of information to {\it all} destination nodes. When source
$n$ has a packet to transmit, it does so with probability $p_n$ in
the first available slot. This scenario is depicted in
Fig.\ref{fig:multicast_NsourceMdest}. Each packet is intended for
all $M$ destinations. We assume that instantaneous and error-free
acknowledgements (ACKs) are sent from the destinations and that each
source-destination pair has a dedicated channel for ACKs. If the
source has not yet received an ACK from all $M$ destinations, the
packet is retransmitted. This policy of relentless retransmissions
is assumed throughout the present work. We note that this policy is
sub-optimal in terms of stable throughput. For instance, in the case
of a single destination $M=1$, collision resolution algorithms such
as the one in \cite{Cape79} have been shown to provide a higher
stable throughput in the infinite-user case. We choose to focus our
attention on random access with retransmissions as a first,
non-trivial step in investigating the stability of random access
broadcast.

\subsection{Queueing stability}
%=========================================

Let $Q_n(k)$ denote the length of the queue at the $n^{th}$ source
node at the beginning of the $k^{th}$ slot in system ${\cal S}$. The
evolution of the queue is expressed as follows.
\begin{equation}
Q_n(k+1) = \left( Q_n(k) - B_n(k) \right)^{+} + A_n(k),
\end{equation}
where $x^+ = x$ if $x \geq 0$, and 0 otherwise. In the above
equation, $A_n(k)$ denotes the arrivals to source $n$, where
$E[A_n(k)] = \lambda_n, \forall k$ and $B_n(k)$ denotes completed
services (or departures) from $n$. Thus $B_n(k)$ takes value 1 if a
packet from source $n$ completes service in slot $k$. We introduce
the service rate $\mu_n \triangleq \lim_{k \rightarrow \infty}
\mbox{Pr} \{ B_n(k)\}$ as the probability that a packet completes
service in the steady-state.

The vector of queue lengths forms an $N$-dimensional, irreducible,
aperiodic Markov chain $\mathbf{Q}(k) = ( Q_1(k), Q_2(k),\hdots,
Q_N(k) )$. The system is stable if, for $\mathbf{x} \in
\mathbb{N}^N$
\begin{equation}
\lim_{\mathbf{x} \rightarrow \infty} \lim_{k \rightarrow \infty}
\mbox{Pr} \{ \mathbf{Q}(k)< \mathbf{x} \} = 1.
\end{equation}
For our Markov chain $\mathbf{Q}(k)$, stability is equivalent  to
positive recurrence of the Markov chain. We define the
\emph{stability region} of the system is the set of all arrival
rates $(\lambda_1, \lambda_2, \hdots, \lambda_N )$ for which there
exists a set of transmission probabilities $( p_1, p_2, \hdots, p_N
)$ such that the system is stable. A primary tool used in our work
is Loynes' result \cite{Loynes62}, which tells us that if $A_n(k)$
and $B_n(k)$ are nonnegative, finite, and strictly stationary, then
source $n$ is stable if and only if $\lambda_n < \mu_n$.

\subsection{Dominant systems}

In the original system ${\cal S}$, we cannot easily write down the
average service rate $\mu_n$ of a source because the service rate
varies depending on whether the other sources are empty or
backlogged. Instead, we introduce a dominant system ${\cal S}^{[1]}$
which behaves exactly like system ${\cal S}$ except that all $N$
sources continue to transmit ``dummy'' packets when empty. The dummy
packets do not affect the information-carrying ability of the
source, but their transmission results in a decoupling of the
queues. In the dominant system ${\cal S}^{[1]}$, all sources behave
as if they are backlogged, the probability of interference from
other sources is known according to the $p_n$ values, and we can
easily write down the service rates $\mu_{n}^{[1]}$. Let $Q_n^{[1]}$
denote the length of the queue at source $n$ in system ${\cal
S}^{[1]}$. It can be shown \cite{TsybakovMikhailov} that if
$\mu_{n}^{[1]} \leq \mu_n$ then $\forall x \in\mathbb{N}$,
\begin{equation*}
\mbox{Pr} \{ Q_n^{[1]} > x \} \geq \mbox{Pr} \{ Q_n > x \}.
\end{equation*}
In other words, the length of the queue in ${\cal S}^{[1]}$ is never
shorter than in ${\cal S}$. So if we find the conditions for
stability in ${\cal S}^{[1]}$, then stability in ${\cal S}$ is
implied. Thus, stability in the dominant system is a sufficient
condition for stability in the original system.

\subsection{Throughput Region}

We define the \emph{throughput region} of the system as the
stability region under the assumption that all sources are always
backlogged. As such, we ignore the burstiness of the arrival
process. The throughput region of the system ${\cal S}$ is
equivalent to the stability region of the dominant system ${\cal
S}^{[1]}$. As we noted above, stability in the dominant system
${\cal S}^{[1]}$ implies stability in the original system ${\cal
S}$, so the throughput region of the system provides an inner bound
to the stability region. It has been conjectured (see \cite{LuoAE06}
and references therein) that the stability region and the throughput
region coincide for $N$ sources and a single destination. This
conjecture suggests that the zero state $Q_n = 0$ plays no role in
determining the ergodicity of the Markov chain $\mathbf{Q}(k)$. A
proof of this conjecture has not been found, and is hindered by the
fact that the stability region for $N$ sources has not been found.

%=========================================================
\section{A network of $N$=2 Sources and $M$=2 Destinations}
\label{section:2x2}
%=========================================================

\subsection{The channel model and service rates}
%==============================================

We introduce a channel model similar to the one in
\cite{NawareEtAl03} to represent the probabilistic reception and MPR
that can be attributed to a wireless channel. We define the
reception probability when a single source transmits as follows. For
$n,m{=}1,2$,
\begin{equation*}
q_{n|n}^{(m)} = \mbox{Pr} \{ \mbox{packet from source $n$ is received at destination $m$ } | \mbox{ only source $n$ transmits}\}.
\end{equation*}
Additionally, for the $N$=2 scenario, we will include the
possibility of capture or MPR with the following reception
probability.
\begin{equation*}
q_{n|1,2}^{(m)} = \mbox{Pr} \{ \mbox{packet from source $n$ is
received at destination $m$ } | \mbox{ both sources transmit}\}.
\end{equation*}
If both $s_1$ and $s_2$ are backlogged, the probability that a
packet transmitted from $s_1$ is successfully received at both
destinations is given by
\begin{equation}
\tau_1= \overline{p_2}q_{1|1}^{(1)}q_{1|1}^{(2)} + p_2q_{1|1,2}^{(1)}q_{1|1,2}^{(2)}
\label{eqn:bothrx}
\end{equation}
where $\overline{p_2}=1-p_2$. In general, $\tau_n, n{=}1,2$ is the
probability that a packet from $s_n$ is received at both
destinations when that source attempts transmission. Similarly, we
define $\phi_n$ as the probability that a successful reception
occurs at $d^{(1)}$ given that $s_n$ transmits and $\sigma_n$ as the
probability that $d^{(2)}$ successfully receives when $s_n$
transmits. When both sources are backlogged, $\phi_1$ and $\sigma_1$
are given by
\begin{eqnarray}
\phi_1 & = & \overline{p_2}q_{1|1}^{(1)}{+}p_2q_{1|1,2}^{(1)} \\
\sigma_1 & = & \overline{p_2}q_{1|1}^{(2)}{+}p_2q_{1|1,2}^{(2)}.
\end{eqnarray}

\begin{figure}
\centering
\includegraphics[width=2.5in]{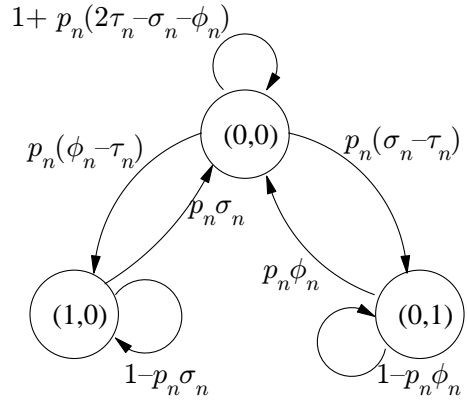}
\caption{The receiver state Markov chain and transition
probabilities for $N=2$ and $M=2$.} \label{fig:recMC}
\end{figure}

We find the service rates $\mu_{1b}$ and $\mu_{2b}$ for the system
in which both sources are backlogged by taking the expected value of
$B_n$. We first condition $B_n$ on the {\it receiver state}, where
$\mathbf{r_n}=(r_n^{(1)}, r_n^{(2)})$ is a vector of binary values
indicating whether the packet currently being transmitted by $s_n$
has been received at each destination. The possible receiver states
are $\mathbf{r_n}=\{ (0,0), (1,0), (0,1)\}$. We do not allow for the
state $(1,1)$ since upon reaching that state, the source will
immediately begin serving the next packet. We can write the
conditional distribution of $B_n$ as given below.
\begin{equation}
\mbox{Pr} \{B_n=1|\bf{r_n}\} =
\begin{cases}
p_n\tau_n, & \text{$\mathbf{r_n}$=(0,0)} \\
p_n\phi_n, & \text{$\mathbf{r_n}$=(0,1)} \\
p_n\sigma_n, & \text{$\mathbf{r_n}$=(1,0)}.
\end{cases}
\label{eqn:B1service}
\end{equation}
To find the expected value of $B_n$ we must first find the
steady-state probability of each receiver state. The set of receiver
states can be modeled by the Markov chain with transition
probabilities depicted in Fig. \ref{fig:recMC}. Let $\pi_{0,0}$,
$\pi_{0,1}$, and $\pi_{1,0}$ denote the steady-state probabilities
of $\bf{r_n}$ when both sources are backlogged. Through use of the
balance equations for the Markov chain and the equation $\pi_{0,0} +
\pi_{0,1} + \pi_{1,0}=1$ we find the steady-state probabilities to
be
\begin{eqnarray*}
\pi_{0,0} & = & \frac{\phi_n\sigma_n}{(\phi_n+\sigma_n)(\phi_n+\sigma_n-\tau_n)-\phi_n\sigma_n} \\
\pi_{0,1} & = & \frac{\sigma_n(\sigma_n-\tau_n)}{(\phi_n+\sigma_n)(\phi_n+\sigma_n-\tau_n)-\phi_n\sigma_n} \\
\pi_{1,0} & = &
\frac{\phi_n(\phi_n-\tau_n)}{(\phi_n+\sigma_n)(\phi_n+\sigma_n-\tau_n)-\phi_n\sigma_n}.
\end{eqnarray*}
Finally, the backlogged service rates $\mu_{nb}, n{=}1,2$ can be
obtained as $\mu_{nb}=E[B_n]=p_n\tau_n\pi_{0,0} + p_n\phi_n\pi_{0,1}
+ p_n\sigma_n\pi_{1,0}$. After simplification, the backlogged
service rates can be expressed as
\begin{equation}
\mu_{nb}=\frac{p_n\phi_n\sigma_n(\phi_n+\sigma_n-\tau_n)}{(\phi_n+\sigma_n)(\phi_n+\sigma_n-\tau_n)-\phi_n\sigma_n}.
\label{eqn:mu_nb}
\end{equation}
We let $\mu_{1e}$ denote the service rate of $s_1$ when $s_2$ is
empty and similarly, $\mu_{2e}$ is the service rate of $s_2$ when
$s_1$ is empty. These service rates can be found directly from the
backlogged service rates as
\begin{equation}
\mu_{1e}= \Bigl. \mu_{1b} \Bigr|_{p_2=0}, \quad \mu_{2e}= \Bigl.
\mu_{2b} \Bigr|_{p_1=0}. \label{eqn:mu_ne}
\end{equation}

%\label{eqn:mu_ne}

\subsection{Stability and throughput regions}
%======================================

We apply the approach introduced in \cite{RaoAE88} to find the
stability region for $N$=2 and $M$=2. For fixed $(p_1, p_2)$ the
following theorem provides the condition for stability.
\begin{theorem} \label{thrm:2x2stabreg_pfix} For a network with $N=2$ sources, $M=2$ destinations,
fixed $(p_1, p_2)$ and $\mu_{nb} \leq \mu_{ne}, n=1,2$, a necessary
and sufficient condition for stability is that the arrival rates lie
within the union of the following two regions.
\begin{equation}
\lambda_1 < \frac{\lambda_2}{\mu_{2b}} \mu_{1b} + \left( 1 -
\frac{\lambda_2}{\mu_{2b}}\right) \mu_{1e}, \quad \lambda_2 <
\mu_{2b}. \label{eqn:S11cond}
\end{equation}
\begin{equation}
\lambda_1  <  \mu_{1b},  \quad \lambda_2  <
\frac{\lambda_1}{\mu_{1b}} \mu_{2b} + \left( 1 -
\frac{\lambda_1}{\mu_{1b}} \right) \mu_{2e}. \label{eqn:S21cond}
\end{equation}
\hspace{0.1cm} %just to add some space before next line
\end{theorem}
This theorem is a generalization of the result in \cite{RaoAE88} and
the proof follows the one provided in that work. To obtain the
stability region over all $(p_1, p_2)$ we formulate a constrained
optimization problem in which we fix $\lambda_1$ and maximize
$\lambda_2$ over $(p_1, p_2)$ subject to Eqns. \ref{eqn:S11cond} and
\ref{eqn:S21cond} \cite{NawareEtAl03}.

Analyzing the throughput region of the system is equivalent to
examining a single dominant system, ${\cal S}^{[1]}$, and the
corresponding condition for stability over all $(p_1, p_2)$. The
stability condition for fixed $(p_1, p_2)$ is expressed as
\begin{equation}
\lambda_1 < \mu_{1b}, \quad \lambda_2 < \mu_{2b}.
\label{eqn:S2cond}
\end{equation}
We can find points on the boundary of the throughput region by again
fixing the value of $\lambda_1$ and maximizing $\lambda_2$ over
$(p_1,p_2)$ subject to Eqn. \ref{eqn:S2cond}. The following theorem
states that the throughput region coincides with the stability
region; a proof is found in the appendix.

\begin{theorem} \label{thrm:2x2Coincide} For a random access broadcast system with $N=2$
source nodes, $M=2$ destination nodes, and service rates as given in
(\ref{eqn:mu_nb}) and (\ref{eqn:mu_ne}), the throughput region is
identical to the stability region.
\end{theorem}

\subsection{Numerical results}
%===================

\begin{table}
\renewcommand{\arraystretch}{1.3}
\caption{The reception probabilities for results shown in Fig.
\ref{fig:results2x2}} \label{table_channels2x2} \centering
\begin{tabular}{|c|c|c|c|c|c|c|c|c|}
\hline
Channel & $q_{1|1}^{(1)}$ & $q_{1|1}^{(2)}$ & $q_{1|1,2}^{(1)}$ &  $q_{1|1,2}^{(2)}$ & $q_{2|2}^{(2)}$ & $q_{2|2}^{(1)}$ & $q_{2|1,2}^{(2)}$ & $q_{2|1,2}^{(1)}$ \\
\hline I & 0.8 & 0.6 & 0.1 & 0.05 & 0.7 & 0.5 & 0.25 & 0.05 \\
\hline II & 0.8 & 0.6 & 0.5 & 0.4 & 0.8 & 0.6 & 0.5 & 0.4 \\
\hline
\end{tabular}
\end{table}

\begin{figure}
\centering
\includegraphics[width=3in]{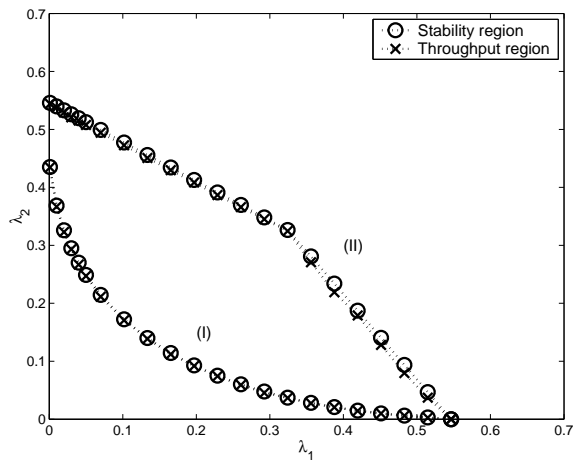}
\caption{Stability and throughput regions for $N$=2 sources and
$M$=2 destinations and two different channel models. The reception
probabilities for channels (I) and (II) are shown in Table
\ref{table_channels2x2}.
 } \label{fig:results2x2}
\end{figure}

In Fig. \ref{fig:results2x2} we show the stability and throughput
regions computed for two different channels with reception
probabilities given in Table \ref{table_channels2x2}. The figure
demonstrates that as the probability of capture or MPR increases,
the stability region transforms from a strictly concave region to a
convex region bounded by straight lines, which agrees with the
results in \cite{NawareEtAl03} for unicast transmission. These
results also demonstrate that for a network with $N$=2 sources and
$M$=2 destinations, the broadcast stability and throughput regions
coincide, confirming Theorem \ref{thrm:2x2Coincide}.

%=================================================
\section{A network of $N$ sources and $M$ destinations}
\label{section:NxM}
%=================================================

\subsection{The channel model and service rates}
%==============================================

The channel model is now a simplified version of the model presented
for the $N$=2, $M$=2 scenario. We assume that whenever two or more
sources transmit simultaneously, none of the transmissions are
successful. Additionally, we assume that the channel reception
probabilities from a source are the same for all destinations,
$q_{n|n}^{(1)}=q_{n|n}^{(2)}= \hdots = q_{n|n}^{(M)}=q_{n|n}$, $n=1
\hdots N$. We refer to the destinations as being {\it
indistinguishable} in this channel model.

We define the receiver-state variable $r_n$ for each source
$n=1,2,\hdots, N$ as the number of destinations that have received
the packet that source $n$ is attempting to transmit, $r_n \in
[0,1,\hdots,M-1]$. We do not allow $r_n$ to take value $M$ since as
soon as all destinations have received the packet, the source will
instantaneously revert back to $r_n = 0$ and either begin serving
the next packet in the queue of source $n$ or become idle if the
source is empty. We define the set ${\cal B}$ as the set of all
sources that are backlogged at the time source $n$ is attempting to
transmit the packet at the front of its queue. Let $\beta_n$ denote
that probability that source $n$ accesses the channel without
interference, i.e.,
\begin{equation}
\beta_n = p_n \left( \prod_{l \in {\cal B} \setminus n}
\overline{p_l}\right).
\end{equation}
The service process conditioned on $r_n$ can be described by
\begin{equation}
\mbox{Pr}\{B_n=1 | r_n=r\} = \beta_n {q_{n|n}}^{M-r}.
\end{equation}

\begin{figure}
\centering
\includegraphics[width=5in]{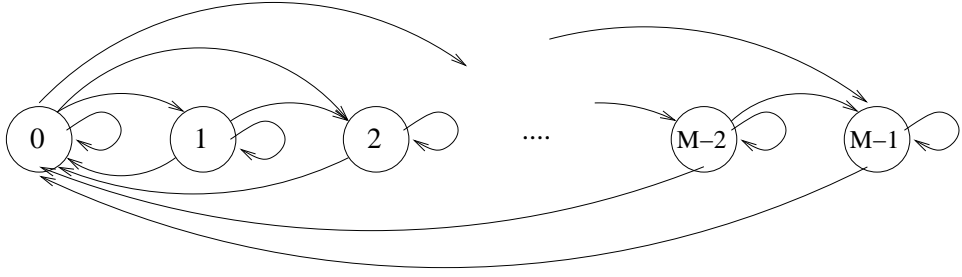}
\caption{Receiver state Markov chain for $M$ indistinguishable
destinations.} \label{fig:MreceiverMC}
\end{figure}

We again develop a Markov chain model for $r_n$ as shown in Fig.
\ref{fig:MreceiverMC}. In this model, transitions ``upward'' can
occur between all pairs of states, however, transitions ``downward''
can only occur between a state and the $0$ state. Additionally, each
state has self-transitions. Let ${\cal P}$ denote the transition
matrix for this Markov chain. When source $n$ accesses the channel
without collision, which happens with probability $\beta_n$, the
transition probability matrix will be a matrix ${\cal P^*}$ which
depends only on the reception probabilities $q_{n|n}$. Otherwise, a
self-transition occurs, corresponding to a transition probability
matrix equal to the identity matrix ${\cal I}$. Thus, we can
describe ${\cal P}$ as the convex combination of two probability
matrices,
\begin{equation}
{\cal P} = \beta_n {\cal P^*} + (1-\beta_n) {\cal I}.
\end{equation}

Let $\pibold$ be the stationary distribution of ${\cal P^*}$,
$\pibold = \pibold {\cal P^*}$. Clearly $\pibold$ will also be the
stationary distribution of ${\cal P}$ since $\pibold {\cal P} =
\beta_n \pibold + (1-\beta_n) \pibold$. We can solve for $\pibold$
as follows. Let $p_{i,j}^{*}$ denote the probability of transition
in $r_n$ from $i$ to $j$ conditioned on $s_n$ accessing the channel
without collision. The transition probabilities are given as
\begin{equation}
p_{i,0}^* =
\begin{cases}
(1-q_{n|n})^M + q_{n|n}^M, & i=0 \\
q_{n|n}^{M-i}, & 0 < i < M.
\end{cases}
\end{equation}
\begin{equation}
p_{i,j}^* =
\begin{cases}
(1-q_{n|n})^{M-i},  & i=j \\
{M-i \choose j-i} q_{n|n}^{j-i} (1-q_{n|n})^{M-j}, & i < j \\
0, & i > j.
\end{cases}
\end{equation}
In order to satisfy $\pi_i = \sum_{j} \pi_j p_{i,j}^{*}$ we have
\begin{eqnarray}
\pi_i & = & \sum_{j=0}^{i-1} \pi_j p_{i,j}^{*} + \pi_i p_{i,i}^{*}
\quad i=1,2,\hdots,M-1 \nonumber\\
 & = & \frac{\sum_{j=0}^{i-1} \pi_j p_{i,j}^{*}}{1-p_{i,i}^{*}}
 \nonumber \\
 & = & \frac{\sum_{k=1}^{i} \pi_{i-k} p_{i-k,i}^{*}}{1-p_{i,i}^{*}}.
\end{eqnarray}
Together with $\sum_{i=0}^{M-1} \pi_i = 1$, we can find the
steady-state probabilities $\pibold$, which is the stationary
distribution of ${\cal P}$.

Once the steady-state probabilities of the receiver Markov chain are
found, the average service rate is given as follows.
\begin{equation}
\mu_n = \beta_n \sum_{i=0}^{M-1} \pi_i  {q_{n|n}}^{M-i}
\end{equation}
We further define
\begin{equation}
\alpha_n = \sum_{i=0}^{M-1} \pi_i  {q_{n|n}}^{M-i}.
\end{equation}
Thus the service rate can be represented in simplified form as
\begin{equation}
\mu_n = \beta_n \alpha_n. \label{eqn:mun_genform}
\end{equation}
This equation summarizes the natural relation between our broadcast
problem and the unicast collision channel problem \cite{LuoAE99}.
The probability that source $n$ completes transmission of a packet,
given by $\mu_n$, is equal to $\beta_n$, the probability that the
source access the channel without collision, times $\alpha_n$, which
is the probability that all destinations receive the packet
conditioned on collision-free access to the channel. In the unicast
collision channel problem, we have $q_{n|n}=1$ and $\alpha_n=1$.
Thus for $\alpha_n=1$ we would expect the stability region for our
broadcast problem to coincide with the results in \cite{LuoAE99} for
the unicast collision channel. This is indeed the case as shown in
the following section.

We define the empty and backlogged service rates as follows. When
${\cal B}$ contains all $N$ sources, the broadcast service rate will
take its minimum value $\mu_{nb}$ where
\begin{equation}
\mu_{nb} = p_n \left( \prod_{l \neq n} \overline{p_l}\right)
\alpha_n. \label{eqn:munb_Mdest}
\end{equation}
The maximum value of $\mu_{n}$ is attained when only source $n$ is
backlogged and all other sources are empty. We denote this service
rate by $\mu_{ne}$
\begin{equation}
\mu_{ne} = p_n  \alpha_n.
\label{eqn:mune_Mdest}
\end{equation}
In deriving bounds on the stability region, we will  take advantage
of the form of $\mu_n$ as expressed in Eqn. \ref{eqn:mun_genform}
and the fact that $\mu_{nb} \leq \mu_n \leq \mu_{ne}$.

\begin{figure}
\centering
\includegraphics[width=3in]{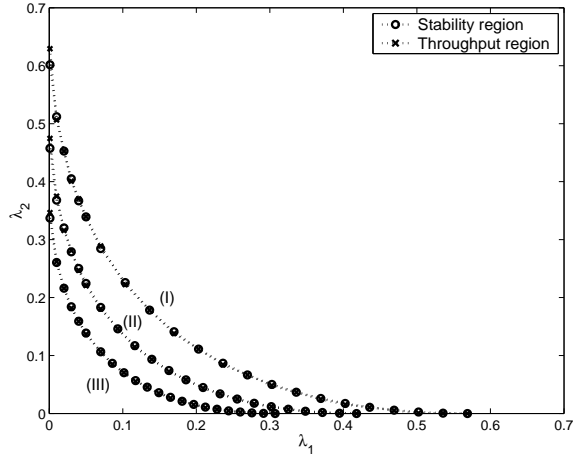}
\caption{Stability and throughput regions for $N$=2 sources and  (I)
2, (II) 5, and (III) 15 destinations.The reception probabilities
are: $q_{1|1}=0.7$, $q_{2|2}=0.8$.} \label{fig:results2xM}
\end{figure}

Before continuing, we observe the effect of the number of
destinations on the stability and throughput regions as shown in
Fig. \ref{fig:results2xM}. These results are for $N$=2 sources and
$M=2,5,$ and 15 destinations. The results are generated using the
approach outlined for $N$=2, $M$=2 with the exception that we use
the backlogged and empty service rates for $M$ destinations as given
in Eqns. \ref{eqn:munb_Mdest} and \ref{eqn:mune_Mdest}. The results
demonstrate that as the number of destinations increases, the
broadcast stability and throughput regions diminish in size.
Additionally, we observe that the stability and throughput regions
coincide. This result is not surprising since the proof of Theorem
\ref{thrm:2x2Coincide} holds for arbitrary $M$ with the channel
model described above.

\subsection{Stability and throughput regions}
%==========================================================

We follow the methodology outlined in \cite{LuoAE99} to develop
bounds on the broadcast stability region. These results are a
generalization of the bounds on the unicast stability region given
in \cite{LuoAE99}. To begin, by the dominant systems argument and
Loynes' result, we can develop loose inner and outer bounds on the
stability region. First, if $\lambda_n < \mu_{nb}$, then since
$\mu_{nb} \leq \mu_n$, queue $n$ must be stable. Furthermore, if
$\lambda_n < \mu_{nb}$ for all $n$, then the entire system is
stable. Likewise, if $\lambda_n > \mu_{nb}$ for all $n$, then the
system is unstable. This follows since $\lambda_n > \mu_{nb}$
corresponds to instability of all the queues in dominant system
${\cal S}^{[1]}$, in which case all of the queues grow to infinity
and the dominant system ${\cal S}^{[1]}$ becomes indistinguishable
from the original system ${\cal S}$ \cite{RaoAE88}. In order to
improve upon these loose bounds, we make use of the {\it stability
rank} of the queues as introduced in \cite{LuoAE99}. Let ${\cal
S}^{[k]}$ denote a dominant system in which sources $k,k+1,\hdots,N$
transmit dummy packets when empty while sources $1,2,\hdots, k-1$ do
not. The proof for the following theorem is the same as in the
original \cite{LuoAE99} with the exception of the constant
$\alpha_n$.

\begin{theorem} \label{thrm:stabrank} Given $\lamNbold$ and $\bf{p_N}$, we order the indices of the sources so that
\begin{equation}
\frac{\lambda_1 (1-p_1)}{\alpha_1 p_1} \leq \hdots \leq
\frac{\lambda_N (1-p_N)}{\alpha_N p_N}. \label{eqn:rank_cond_allN}
\end{equation}
Then in system ${\cal S}$ and any dominant system ${\cal S}^{[k]}$,
if queue $i$ is stable and $j < i$, then queue $j$ is also stable.
\end{theorem}

We can find conditions for the stability of the system through the
procedure outlined in \cite{LuoAE99}. After ordering the sources
according to stability rank, we first check for stability of queue 1
in system ${\cal S}^{[1]}$. If we find that queue 1 is unstable, we
can conclude that the entire system is unstable. However, if queue 1
is stable, we proceed by examining queue 2 in system ${\cal
S}^{[2]}$. Queue 1 is known to be stable in ${\cal S}^{[2]}$ due to
the stability rank. Given stability of queue 1, we will check
whether queue 2 is stable in ${\cal S}^{[2]}$. The procedure
continues in which the stability of queue $k$ in system ${\cal
S}^{[k]}$ is verified assuming that queues $1,2,\hdots,k-1$ are all
stable in ${\cal S}^{[k]}$. If we can finally conclude that queue
$N$ is stable in system ${\cal S}^{[N]}$, then this implies that the
original system is stable.

%We turn our attention now to determining whether source $k$ is
%stable in system ${\cal S}^{[k]}$ for $1 \leq k \leq N$. According
%to the procedure outlined above, we can assume that the lower-ranked
%sources $1,2,\hdots, k-1$ are all stable. To determine whether
%source $k$ is stable, we will develop bounds on the average service
%rate $\mu_k^{[k]}$ for source $k$ in system ${\cal S}^{[k]}$. We can
%express $\mu_k^{[k]}$ as

Assuming that sources $1,2,\hdots, k-1$ are all stable, we will
develop bounds on the average service rate $\mu_k^{[k]}$ for source
$k$ in system ${\cal S}^{[k]}$ in order to help determine whether
$s_k$ is stable. We can express $\mu_k^{[k]}$ as
\begin{equation}
\mu_k^{[k]} = \alpha_k p_k P_E^{[k]} \prod_{j=k+1}^{N} (1-p_j),
\label{eqn:mukk_form}
\end{equation}
where $P_E^{[k]}$ is the probability that none of the sources
$1,2,\hdots,k-1$ transmit in the dominant system ${\cal S}^{[k]}$.
One way to bound $\mu_k^{[k]}$ is by bounding $P_E^{[k]}$ when
expressed as
\begin{multline}
P_E^{[k]} = 1 - Pr\{\mbox{only one of sources $1,2,\hdots,k-1$
transmits}\} \\ - Pr\{\mbox{more than one of sources
$1,2,\hdots,k-1$ transmit}\}. \label{eqn:PEexpr}
\end{multline}
The bounds on $\mu_k^{[k]}$ result in the following two theorems.

\begin{theorem} \label{thrm:suffcond} {\it Sufficient condition.} Given an $N$ source, $M$ destination random access system with $\lamNbold$, $\bf{p_N}$ and the sources ordered according to the stability rank as in Eqn. \ref{eqn:rank_cond_allN}, if $\forall k$, $1 \leq k \leq N$, $\lambda_k < B_k$, where $B_k$ is defined below, then the system is stable.
\begin{eqnarray}
B_k & = & max(C_k, D_k) \nonumber \\
B_1 & = & \alpha_1 p_1 \prod_{j=2}^{N}(1-p_j) \\
C_k & = & \frac{\alpha_k p_k}{1-p_k} \left[ \prod_{j=k}^{N} (1-p_j) - \frac{\sum_{i=1}^{k-1}\lambda_i}{\min\limits_{1 \leq l \leq k-1}\alpha_l} - \frac{1}{2} \sum_{j=1}^{k-1} \left( \frac{\lambda_j p_j}{B_j} \prod_{i=k}^{N}(1-p_i) -\frac{\lambda_j}{\alpha_j} \right) \right] \nonumber \\
D_k & = & \frac{\alpha_k p_k}{1-p_k} \prod_{j=1}^{N}(1-p_j) \left( 1 + \sum_{i=1}^{k-1} \left( 1- \frac{\lambda_i}{B_i} \right) \frac{p_i}{1-p_i} \right)\nonumber \\
\end{eqnarray}
\end{theorem}

\begin{theorem} \label{thrm:nececond}
{\it Necessary condition.} Given an $N$ source, $M$ destination
random access system with $\lamNbold$, $\bf{p_N}$ and the sources
ordered according to the stability rank as in Eqn.
\ref{eqn:rank_cond_allN} a necessary condition for stability of the
system is that $\forall k$,
\begin{equation}
\lambda_k \leq \frac{\alpha_k p_k}{1-p_k} \left(
\prod_{j=k}^{N}(1-p_j) -
\frac{\sum_{i=1}^{k-1}\lambda_i}{\max\limits_{1 \leq l \leq
k-1}\alpha_l} \right).
\end{equation}
\hspace{0.1cm} %just to add some space before the next line
\end{theorem}

The proofs for Theorems \ref{thrm:suffcond} and \ref{thrm:nececond}
are described in the Appendix. We make use of these two theorems to
develop bounds on the stability region by fixing the values of
$\lambda_1,\lambda_2,\hdots,\lambda_{N-1}$ and numerically
optimizing $\lambda_N$ over all $\mathbf{p_N} \in [0,1]^N$ subject
to the inequalities given in the Theorems.

The throughput region of the network of arbitrary size can be
determined exactly using the backlogged service rates in Eqn.
\ref{eqn:munb_Mdest}. Again, we fix $\lambda_1, \lambda_2, \hdots,
\lambda_{N-1}$ and maximize $\lambda_N$ over all $\bf{p_N}$ subject
to $\lambda_N < \mu_{Nb}$.

\subsection{Numerical results}
%==========================================

\begin{table}
\renewcommand{\arraystretch}{1.3}
\caption{Optimization results of stability bounds and throughput for
a network of $N$=4 sources and $M$=8 and 10 destinations. The values
of $\lambda_1, \lambda_2,$ and $\lambda_3$ were fixed and
$\lambda_4$ has been optimized over all $\mathbf{p}$.}
\label{table:NxMresults1}
\begin{center}
\begin{tabular}{|c|c|c|c|c|c|c|c|c|c|c|}
\hline
$M$ & $q_{1|1}$ & $q_{2|2}$ & $q_{3|3}$ & $q_{4|4}$ & $\lambda_1$ & $\lambda_2$ & $\lambda_3$ & Stability-upper & Stability-lower & Throughput\\
\hline
8 & 0.9 & 0.8 & 0.7 & 0.9 & 0.01 & 0.01 & 0.01 & 0.3648 & 0.3213 & 0.3213 \\
  &     &     &     &     & 0.07 & 0.02 & 0.01 & 0.2125 & 0.1672 &
  0.1672 \\
  &     &     &     &     & 0.05 & 0.05 & 0.05 & 0.1363 & 0.0566 &
  0.0566 \\
  &     &     &     &     & 0.07 & 0.05 & 0.05 & 0.1090 & 0.0376 &
  0.0376 \\ \hline
8 & 0.8 & 0.8 & 0.8 & 0.8 & 0.01 & 0.01 & 0.01 & 0.2527 & 0.2433 &
0.2434 \\
  &     &     &     &     & 0.07 & 0.02 & 0.01 & 0.1294 & 0.1090 &
  0.1090 \\
  &     &     &     &     & 0.05 & 0.05 & 0.05 & 0.0784 & 0.0428 &
  0.0428 \\
  &     &     &     &     & 0.07 & 0.05 & 0.05 & 0.0587 & 0.0254 &
  0.0254 \\ \hline
10 & 0.8 & 0.8 & 0.8 & 0.8 & 0.01 & 0.01 & 0.01 & 0.2329 & 0.2236 &
0.2236 \\
   &     &     &     &     & 0.07 & 0.02 & 0.01 & 0.1153 & 0.0951 &
   0.0951 \\
   &     &     &     &     & 0.05 & 0.05 & 0.05 & 0.0651 & 0.0318 &
   0.0321 \\
   &     &     &     &     & 0.065 & 0.05 & 0.05 & 0.0503 & 0.0196 &
   0.0196 \\ \hline
\end{tabular}
\end{center}
\end{table}

\begin{table}
\renewcommand{\arraystretch}{1.3}
\caption{Optimization results of stability bounds and throughput for
a network of $N$=5 and 10 sources and $M$=10 destinations. The
reception probabilities are $q_{n|n}=0.8$ for $n=1,\hdots,N$. The
values of $\lambda_1, \hdots, \lambda_{N-1}$ were fixed and
$\lambda_N$ has been optimized over all $\mathbf{p}$.}
\label{table:NxMresults2}
\begin{center}
\begin{tabular}{|c|c|c|c|c|}
\hline
$N$ & $\lambda_1, \hdots, \lambda_{N-1}$ & Stability-upper & Stability-lower & Throughput\\
\hline

5 & $[0.010, 0.010, 0.010, 0.010]$ & 0.2078 & 0.1939 & 0.1939 \\
  & $[0.070, 0.020, 0.010, 0.010]$ & 0.1051 & 0.0789 & 0.0789 \\
  & $[0.035, 0.035, 0.035, 0.035]$ & 0.0751 & 0.0362 & 0.0362 \\
  & $[0.050, 0.035, 0.035, 0.035]$ & 0.0602 & 0.0223 & 0.0223 \\ \hline
10 & $[0.010, 0.010, \hdots 0.010]$ & 0.1266 & 0.0912 & 0.0912 \\
   & $[0.070, 0.010, \hdots 0.010]$ & 0.0679 & 0.0252 & 0.0252 \\
   & $[0.017, 0.017, \hdots 0.017]$ & 0.0621 & 0.0137 & 0.0137 \\
   & $[0.020, 0.017, \hdots 0.017]$ & 0.0591 & 0.0108 & 0.0108 \\
   \hline
\end{tabular}
\end{center}
\end{table}

A collection of results on stability and throughput for various
values of $N$ and $M$ are shown in Tables \ref{table:NxMresults1}
and \ref{table:NxMresults2}. The values in Table
\ref{table:NxMresults1} demonstrate the effect of the channel
reception probabilities and the number of destinations on the
stability and throughput regions. The trends in these results are
the same as those observed for a network of $N$=2 sources. In Table
\ref{table:NxMresults2} we observe that as the number of sources $N$
increases, the stability and throughput regions diminish in size.
Furthermore, in all cases, the throughput values fall within the
upper and lower bounds on the stability values. As such, these
results support the conjecture that the stability and throughput
regions coincide. Of special note, the lower bound for stability and
the throughput value appear to be equal in many cases. This is not
entirely true. In the results shown here, the throughput value is in
fact slightly larger than the lower bound on the stability value,
but the difference is at most $10^{-6}$.

%===================
\section{Conclusions} \label{section:conclusion}
%===================

In this work we have investigated the information-carrying ability
of source nodes which compete for access to a shared channel in
order to broadcast messages to a common set of destination nodes.
The scenario we consider is relevant to broadcast applications and
network discovery and control. We formulated the problem as a
network of queues which interact through the shared channel and
investigated the stability and throughput regions of the source
nodes. Our results strengthen an unproven conjecture that the
stability and throughput regions of finite-user random access
coincide. We also demonstrate that the throughput and stable
throughput regions diminish as the number of source or destination
nodes grows. Our intention in this work is to present results
relating to the scaling laws for wireless networks, which often show
a degradation in throughput as the number of nodes increases. Our
results show a similar trend from a multicast stability point of
view.

\section{Appendix}

\subsection{Proof of Theorem \ref{thrm:2x2Coincide}}

We show that the boundaries of the stability and throughput regions,
given by the results of two constrained optimization problems, are
identical. We replace $\lambda_1$ by $x$ and $\lambda_2$ by $y$.
Furthermore, we express the backlogged service rates $\mu_{nb}$ in
Eqn. \ref{eqn:mu_nb} as follows.
\begin{equation}
\mu_{1b} = p_1 f_1(p_2), \quad \mu_{2b} = p_2 f_2(p_1).
\end{equation}
The empty service rates can then be expressed as $\mu_{1e} = p_1
f_1(0)$ and $\mu_{2e}=p_2 f_2(0)$. In these expressions, $f_1(p_2)$
is functionally independent of $p_1$ and decreasing in $p_2$,
corresponding to $\mu_{1b} \leq \mu_{1e} \iff f_1(p_2) \leq f_1(0)$.
Similarly, $f_2(p_1)$ is independent of $p_2$ and decreasing in
$p_1$. With this notation in place, the {\it boundary} of the region
given in Theorem \ref{thrm:2x2stabreg_pfix} for fixed $(p_1,p_2)$
can be written as follows.
\begin{eqnarray}
x = p_1f_1(0) - \frac{y p_1 \left( f_1(0) - f_1(p_2)\right)}{p_2
f_2(p_1)}, \quad 0 \leq y \leq p_2 f_2(p_1) \label{eqn:SxOptim} \\
y = p_2 f_2(0) - \frac{x p_2\left( f_2(0) - f_2(p_1)\right)}{p_1
f_1(p_2)}, \quad 0 \leq x \leq p_1 f_1(p_2) \label{eqn:SyOptim}
\end{eqnarray}
To find the stability region, we should maximize the expressions in
(\ref{eqn:SxOptim}) and (\ref{eqn:SyOptim}) over $(p_1,p_2)$ and
take the intersection of the regions bounded by the resulting
curves. We note that this is not a standard optimization problem
because the objective function is piece-wise and non-differentiable
at a point in its domain. An example of an analytical solution to
this optimization problem is given in \cite{NawareEtAl03} for the
single destination case.

The boundary of the throughput region for fixed $(p_1,p_2)$ as given
in Eqn. \ref{eqn:S2cond} can be written as follows.
\begin{eqnarray}
x = p_1 f_1(p_2), \quad 0 \leq y \leq p_2 f_2(p_1)
\label{eqn:TxOptim} \\
y = p_2 f_2(p_1), \quad 0 \leq x \leq p_1 f_1(p_2)
\label{eqn:TyOptim}
\end{eqnarray}
The throughput region is found by maximizing the expressions in
(\ref{eqn:TxOptim}) and (\ref{eqn:TyOptim}) and taking the
intersection of the resulting regions. Consider Eqn.
(\ref{eqn:TyOptim}) in which we wish to maximize $y$ over
$(p_1,p_2)$. Note that the constraint $x \leq p_1 f_1(p_2)$ is a
lower bound on $p_1$ over which we perform the maximization. Since
$f_2(p_1)$ is decreasing in $p_1$, the lower bound $p_1 \geq
x/f_1(p_2)$ provides an upper bound on $y$, and this upper bound can
be achieved when maximizing (\ref{eqn:TyOptim}) over $(p_1,p_2)$.
Then maximization of $y$ in (\ref{eqn:TyOptim}) is equivalent to
maximizing $y$ as follows.
\begin{equation}
y = \biggl.p_2 f_2(p_1) \biggr|_{p_1 = \tfrac{x}{f_1(p_2)}}, \quad 0
\leq x \leq p_1 f_1(p_2)
\end{equation}
To see that this is identical to (\ref{eqn:SyOptim}), we write
$f_2(p_1)$ as follows.
\begin{eqnarray}
f_2(p_1) & = & f_2(0) - \left( f_2(0) - f_2(p_1) \right) \\
 & = & f_2(0) - \frac{p_1 \left( f_2(0) - f_2(p_1) \right)}{p_1}
\end{eqnarray}
Then
\begin{equation}
\biggl.p_2 f_2(p_1) \biggr|_{p_1 = \tfrac{x}{f_1(p_2)}} = p_2 f_2(0)
- \frac{x p_2\left( f_2(0) - f_2(p_1)\right)}{p_1 f_1(p_2)}
\end{equation}
and the maximum $y$ in (\ref{eqn:TyOptim}) is identical to the
maximum $y$ in (\ref{eqn:SyOptim}). Similarly, the maximum of $x$ in
(\ref{eqn:TxOptim}) is identical to the maximum of $x$ in
(\ref{eqn:SxOptim}).

\subsection{Proof of Theorem \ref{thrm:suffcond}}
Our proof follows the one given in \cite{LuoAE99}. A sufficient
condition for stability corresponds to bounding $\mu_k^{[k]}$ from
below. We obtain two separate lower bounds, $C_k$ and $D_k$, and
take their maximum to obtain the result. The bound $C_k$ is derived
through Eqns. \ref{eqn:mukk_form} and \ref{eqn:PEexpr}. Since
sources $1,2,\hdots, k-1$ are stable, the following holds.
\begin{multline}
Pr\{\mbox{success by one of the sources $1,2,\hdots,k-1$}\}  =  \sum_{i=1}^{k-1} \lambda_i \\
  \geq \left( \min_{1 \leq l \le k} \alpha_l \right) Pr\{\mbox{only one of the sources $1,2,\hdots,k-1$ transmits}\} \prod_{j=k}^{N} (1-p_j)
\end{multline}
This indicates that
\begin{equation}
Pr\{\mbox{only one of the sources $1,2,\hdots,k-1$ transmits}\}
 \leq \frac{\sum_{i=1}^{k-1} \lambda_i}{\left( \min\limits_{1 \leq l \le k} \alpha_l \right) \prod_{j=k}^{N}
 (1-p_j)}.
\label{eqn:PE1_upper}
\end{equation}
For the next term in $P_E^{[k]}$ we have
\begin{multline}
Pr\{\mbox{a packet from $j$ collides with others from $1,2,\hdots,k-1$}\}   \\
 = Pr\{\mbox{$j$ transmits}\} - Pr\{\mbox{$j$ transmits \& no others from $1,2,\hdots,k-1$ transmit}\} \\
 = p_j Pr\{\mbox{$j$ backlogged}\} - \frac{Pr\{\mbox{$j$ transmits \& no others transmit}\}}{\prod_{i=k}^{N}(1-p_i)} \\
 = p_j \frac{\lambda_j}{\mu_j^{[k]}} - \frac{\lambda_j}{\alpha_j
 \prod_{i=k}^{N}(1-p_i)}.
\label{eqn:derive_sumterm}
\end{multline}
We also make use of the result shown in \cite{RaoAE88} that for $j<k$, $\mu_{j}^{[j]} \leq \mu_{j}^{[k]}$. Thus, the second term in $P_E^{[k]}$ is upper bounded as follows.
\begin{equation}
Pr\{\mbox{more than one of the sources $1,2,\hdots,k-1$ transmits}\} \leq \frac{1}{2} \sum_{j=1}^{k-1} \left( p_j
 \frac{\lambda_j}{\mu_j^{[j]}} - \frac{\lambda_j}{\alpha_j \prod_{i=k}^{N}(1-p_i)}  \right)
\label{eqn:PE2_upper}
\end{equation}
By combining Eqns. \ref{eqn:PEexpr}, \ref{eqn:PE1_upper}, and \ref{eqn:PE2_upper} we obtain the lower bound
\begin{equation}
P_E^{[k]} \geq 1 - \frac{\sum_{i=1}^{k-1} \lambda_i}{\left(
\min\limits_{1 \leq l \le k} \alpha_l \right) \prod_{j=k}^{N}
(1-p_j)} - \frac{1}{2} \sum_{j=1}^{k-1} \left( p_j
\frac{\lambda_j}{\mu_j^{[j]}} - \frac{\lambda_j}{\alpha_j
\prod_{i=k}^{N}(1-p_i)}  \right).
\end{equation}
Together with Eqn. \ref{eqn:mukk_form}, this provides $C_k$, our first lower bound on $\mu_{k}^{[k]}$.

The other lower bound on $\mu_{k}^{[k]}$ is derived using an approach from \cite{RaoAE88}. We have adapted it for the broadcast problem below and refer to it as $D_k$.
\begin{multline}
\mu_k^{[k]} \geq \frac{\alpha_k p_k}{1-p_k} Pr\{\mbox{no other source transmits}\} \\
+ \frac{\alpha_k p_k}{1-p_k} Pr\{\mbox{one of the first $1,2,\hdots,k-1$ sources transmits but is empty}\}  \\
= \frac{\alpha_k p_k}{1-p_k} \prod_{j=1}^{N}(1-p_j) \left( 1 + \sum_{i=1}^{k-1} \left( 1- \frac{\lambda_i}{\mu_i^
{[i]}} \right) \frac{p_i}{1-p_i} \right) = D_k.
\end{multline}
Note that we can express the exact value of $\mu_{1}^{[1]}$ as
$\mu_{1}^{[1]}=\alpha_1 p_1 \prod_{j=2}^{N}(1-p_j)$. By beginning
with $B_1=\mu_{1}^{[1]}$ we can iterate through $k$ values
$2,\hdots,N$ to obtain the result.

The proof of Theorem \ref{thrm:nececond} develops upper bounds on
$\mu_k^{[k]}$ and thus on $P_E^{[k]}$. The technique is similar to
the one outlined above in finding $C_k$.

\section*{Acknowledgment}
The views and conclusions contained in this document are those of
the authors and should not be interpreted as representing the
official policies, either expressed or implied, of the Army Research
Laboratory or the U. S. Government.

% optional entry into table of contents (if used)
%\addcontentsline{toc}{section}{Acknowledgment}


% http://www.ctan.org/tex-archive/macros/latex/contrib/supported/IEEEtran/bibtex
\bibliographystyle{IEEEtran}
\bibliography{IEEEabrv,Correspondence}

\end{document}